\documentclass[%
 reprint,
bibnotes,
 amsmath,amssymb,
 aps,
 superscriptaddress,
prb,
floatfix,
]{revtex4-2}

\usepackage{graphicx}
\usepackage{dcolumn}
\usepackage{bm}
\usepackage{diagbox}
\usepackage{hyperref} 
\usepackage{caption}

\begin{document}

\preprint{APS/123-QED}

\title{Effect of non-homogeneous non-magnetic Impurities in Superconductors:\\
A comparison between Microscopic and Macroscopic theories}%

\author{Carlos Redondo Herrero}
\affiliation{
Université Paris-Saclay, CNRS/IN2P3, IJCLab, Orsay, France
}%
\author{Tejas Guruswamy}
\affiliation{
Argonne National Laboratory, Lemont, IL, USA
}%
\author{Orlando Quaranta}
\affiliation{
Argonne National Laboratory, Lemont, IL, USA
}
\affiliation{University of Chicago, Chicago, IL, USA
}
\author{Akira Miyazaki}%
\affiliation{
Université Paris-Saclay, CNRS/IN2P3, IJCLab, Orsay, France
}%

\date{\today}

\begin{abstract}
Non-homogeneous, non-magnetic impurities in superconductivity have recently gained traction as a promising approach to reduce surface resistance and increase the superheating field. We aim to deepen the understanding of impurities' role from a theoretical perspective by utilizing Eilenberger's equation, and to investigate the validity of a more phenomenological approach based on London's equation for determining the magnetic field profile within the superconductor. 
We show that the microscopic and macroscopic theories produce identical electromagnetic field distributions for any spatial distributions of impurities inside the superconductors.
However, the two approaches provide different magnetic field-related quantities.
The macroscopic model only gives the Bean-Livingston barrier, while the microscopic theory provides the experimentally-relevant superheating field.
Based on the microscopic formalism, we determine a family of impurity profiles that make the superheating field $H_{\rm sh}$ equal to the critical field $H_c$, which is the maximum achievable. 
\end{abstract}

\maketitle


\section{Introduction}
Impurities are an essential part of material science and condensed matter physics, as they intrinsically appear in materials and change their properties. 
Since the periodicity of the crystal is a fundamental property when determining the Brillouin zone, the energy bands, and the Bloch wavefunctions~\cite{Ashcroft:102652}, introducing impurities breaks translational symmetry in the crystal. This makes the study of impurities a fundamental challenge, 
even with the use of a wide range of computational techniques, such as Density Functional Theory~\cite{Giannozzi_2009}.
Both theoretical and experimental studies are crucial to quantitatively understand the role of impurities in condensed matter physics.
In this paper, we shed light on the role of non-magnetic impurities in conventional superconductors under external electromagnetic fields.

In superconductivity, despite Anderson's theorem, non-magnetic impurities play an important role in the surface impedance and in the superheating field. 
When there is an external magnetic field, 
the vector potential can be seen as a Doppler shift that changes the Fermi surface.
Since non-magnetic impurities are proportional to the Fermi surface-averaged Green functions,
this Doppler shift violates an assumption of Anderson's theorem, and the critical temperature can be changed without magnetic impurities.
Such impurities within the penetration depth of the surface manipulate the optical conductivity and thus surface impedance of superconductors~\cite{benvenuti99}.
In type II superconductors, the transition to the normal state occurs only after passing through a metastable vortex state. This transition happens once the applied field exceeds the superheating field, $H_{\rm sh}$, 
a quantity that can itself be altered by the distribution of impurities within the superconductor.
Beyond this field, the metastable regime becomes unstable as the Cooper pairs begin depairing. 
Even for conventional {\it s}-wave superconductors, non-equilibrium behavior under a strong external field with impurity scattering is not fully understood theoretically.

The majority of the literature has been limited to uniformly distributed impurities~\cite{PhysRevB.85.054513, PhysRevB.78.224509} . Inhomogeneously distributed impurities have recently been identified as key to understanding the performance of superconducting resonators, with experimental evidence from Secondary Ion Mass Spectrometry~\cite{10.1063/5.0308624}.
Such a spatial distribution can be engineered by a dedicated process of doping or baking the superconductors to diffuse interstitial atoms within the superconducting penetration depth from the surface.
Two theoretical frameworks were independently proposed to explain the role of spatial inhomogeneity in non-magnetic impurities.
Ngampruetikorn and Sauls~\cite{PhysRevResearch.1.012015} solved the Eilenberger equations for $H_{\rm sh}$ to extend the previous calculation for uniformly distributed impurities~\cite{PhysRevB.85.054513, PhysRevB.78.224509} to disorders exponentially distributed from the surface.
A steep gradient of the impurity profile from the surface enhances the $H_{\rm sh}$.
Lechner et al.~\cite{10.1063/5.0191234} estimated the Bean-Livingston barrier via a phenomenologically {\it modified} London equation with position dependence through a spatial distribution of mean free path.
The latter approach can be considered as a continuous limit of the multi-layer model~\cite{10.1063/1.2162264, Kubo_2014, herrero2026analyticalevaluationsurfacebarrier}.
These microscopic theories and macroscopic models have qualitative agreement; however, quantitative comparison has been lacking.

In this paper, we show the equivalence of the above microscopic and macroscopic approaches and justify the use of the naively modified London equation.
We also reveal the limitations of such a simple method by comparisons to calculations of $H_{\rm sh}$ from the microscopic theory.
The Eilenberger equations are solved for an arbitrary spatial distribution of non-magnetic impurities, and an optimum disorder distribution is proposed.
We also see that the assumption of Anderson's theorem is no longer valid under an external field; thus, even non-magnetic impurities can impact the superconducting gap and critical temperature of {\it s}-wave superconductors.


\section{Theory} \label{sec.Theory}
\subsection{Microscopic Theory}
The original effective Hamiltonian by Bardeen, Cooper, and Schrieffer~\cite{bardeen57} did not include any impurity scattering. 
This scattering is typically included through a random impurity potential. Subsequently, the Green function is modified by calculating the ``self-energy'' at 1-loop,
leading to the Gorkov equation with impurities.
Since the Fermi energy is much larger than the other relevant degrees of freedom \cite{Polchinski:1992ed}, integrating over the energy yields the quasiclassical approximation. 
Finally, one arrives at the Eilenberger equation by imposing the correct boundary conditions for a semi-infinite superconductor in London's gauge:
\begin{equation}
    \left[ \left( i\hbar\omega_n-\frac{1}{2}\hbar \bm{v}_F\cdot \bm{q} \right)\hat{\tau}_3-\hat{\Sigma}_{\text{imp}}+i\hat{\tau}_2\Delta,\,\hat{g} \right]=i\bm{v}_F\cdot\nabla\hat{g}, \label{eq:Eilenberger1}
\end{equation}
\begin{equation}
\hat{g}\cdot \hat{g}=1, \label{eq:Eilenberger2}
\end{equation}
\begin{equation}
    \hat{g}=g\hat{\tau}_3+f\hat{\tau}_2, \label{eq:Eilenberger3}
\end{equation}
\begin{equation}
    \left.\nabla \times\bm{A}\right|_{S}=\mu_0\bm{H}_0, \quad \lim_{r\to\infty}\bm{A}=0, \label{eq:Eilenberger4}
\end{equation}
where the commutation relation
\begin{equation}
    [A, B] = AB - BA,
\end{equation}
is introduced.
Here $\omega_n$ are the fermionic Matsubara frequencies, $\bm{v}_F$ is the Fermi velocity, $\bm{q}$ is the superfluid momentum. 
In London's gauge, $\bm{q}=\left(-2e/\hbar\right)\bm{A}$, with $\bm{A}$ being the electromagnetic vector potential. $\Delta$ is the superconducting gap, $\hat{\Sigma}_{\text{imp}}=-i\gamma\langle \hat{g}\rangle_F$ is the impurity ``self-energy'' which is proportional to the impurity profile $\gamma=\gamma(\bf r)$ and $\hat{g}$ averaged over the Fermi surface, denoted $\langle\cdots\rangle_F$, where $\hat{g}$ is the $2\times2$ quasiclassical Green's function with normal and anomalous components, $\bm{H}_0$ is the external magnetic field, and $\hat{\bm{\tau}}$ are the Pauli matrices. 
In the extreme type II superconductivity limit, the gradient term is negligible due to the Green functions slowly varying in space, as the coherence length is much smaller than the penetration depth $\xi\ll\lambda$.  

To find the unknown superconducting gap, the self-consistent gap equation must be solved, which in the Matsubara space is
\begin{equation}
    \Delta(\bm{r},T)=2\pi k_B Tg\sum_{n=0}^\infty \langle f\rangle_F,\label{eq:gap}
\end{equation}
with $g$ being the coupling of the four-point effective interaction, and the Matsubara sum has a cutoff at the Debye frequency.

We consider the superconducting current to determine the magnetic and electric fields inside the superconductor.
To do so, we impose N\"{o}ether's theorem on a $U(1)$ global symmetry, which is related to probability conservation.
This leads to a four-vector current-conservation equation. After multiplying by the charge and applying the quasiclassical approximation, the superconducting current is obtained.
\begin{equation}
    \pmb{J}(\bm{r},T)=-2\pi eN(0)k_B T \sum_{n=0}^\infty\text{Im}\langle\bm{v}_F g\rangle_F\label{eq:current} .
\end{equation}
Finally, the current is related to the electromagnetic vector potential through Maxwell's equations, as
\begin{equation}
    \bm{J}(\bm{r},T)=\frac{1}{\mu_0}\nabla\times(\nabla\times\bm{A})\label{eq:current_vec} .
\end{equation}
To find the correct solution inside the superconductor, we need to solve these Eq.~\eqref{eq:Eilenberger1}-\eqref{eq:Eilenberger4} and~\eqref{eq:gap}-\eqref{eq:current_vec} self-consistently. 
Ngampruetikorn and Sauls~\cite{PhysRevResearch.1.012015} took the same approach and considered a specific impurity profile
\begin{equation}
    \gamma(x) = \gamma_0 \exp{(-x/\zeta)}.
\end{equation}
In this paper, we consider more general distributions.

\subsection{Macroscopic Model}
The macroscopic models are much simpler than the microscopic one that we just described; however, they ignore much of the underlying physics. 
This is a double-edged sword, as it leads to a much simpler solution at the cost of losing valuable information about the superconducting state.
Out of several macroscopic models, the London equations were originally introduced phenomenologically~\cite{10.1098/rspa.1935.0048} 
and recovered from the microscopic theory by considering the diamagnetic current in the local limit~\cite{bardeen57}.
A spatially uniform non-magnetic impurity scattering is usually included by substituting the London penetration depth for the Pippard one~\cite{tinkham2004introduction}.

To treat spatially inhomogeneous disorders, the above approach was recently generalized to include a spatial distribution of the penetration depth~\cite{10.1063/5.0191234}
\begin{equation}
    \lambda^2\nabla^2\bm{B}-2\lambda\lbrace\nabla_{\bm{B}}[(\nabla\lambda) \cdot \bm{B}]-(\nabla\lambda \cdot \nabla)\bm{B}\rbrace=\bm{B}, \label{eq:modi_Lon}
\end{equation}
where $\lambda$ and $\bm{B}$ have a spatial dependence. $\nabla_{\bm{B}}$ indicates that the derivative only applies to $\bm{B}$. 
Eq.~\eqref{eq:modi_Lon} can be denoted more simply in tensor notation
\begin{equation}
    \sum_{j=1}^3\lambda^2\partial^2_jB_i-2\lambda(\partial_j\lambda)(\partial_i B_j-\partial_j B_i)=B_i.
\end{equation}
The spatial dependence of the penetration depth is included via the mean-free path as
\begin{equation}
    \lambda(\bm{r})=\lambda_0\sqrt{1+\frac{\xi_0}{l(\bm{r})}}, \quad l(\bm{r})=\frac{\pi\xi_0}{2\gamma(\bm{r})}. \label{eq:penetration_depth}
\end{equation}
Including the spatial dependence of the penetration depth was first introduced by Simon and Chaikin~\cite{PhysRevB.23.4463, PhysRevB.30.3750} in analyzing the proximity effect.
Checchin and Grassellino \cite{Checchin:2017kix, osti_1348232} applied Eq.~\eqref{eq:modi_Lon} for superconducting resonator issues,
{\it a priori} assuming the penetration depth is spatially dependent as a complementary error function.
Lechner et al.~\cite{10.1063/5.0191234} were the first to combine Eq.~\eqref{eq:modi_Lon} and \eqref{eq:penetration_depth} together with the oxygen profile near the surface via the diffusion equation.
To the best of our knowledge, this macroscopic model has not been justified from microscopic theories.

In the following sections, we will study the validity of this phenomenological approach. It is useful for obtaining the magnetic field profile.
But since the equations do not take into account the superconducting gap nor the Green functions, we cannot know anything regarding the density of states nor the phase transition to the normal state.  This implies that the superheating fields cannot be obtained using this macroscopic model.
Thus, this approach loses important information despite being used in the literature and validated by this paper.

\section{Numerical calculations} \label{sec:Code}

The code is organized around two levels of self-consistency. At the \emph{local} level, for given $(T, h, \gamma)$, the impurity-renormalized Green functions and the gap $\Delta$ are determined simultaneously. At the \emph{spatial} level, the local solutions at each grid point are coupled through Amp\`ere's equation, which updates the vector potential $A(x)$ and hence the Doppler shift $h(x)$.

For a single spatial point, the local (uniform) solve proceeds inward from the Matsubara sum: at each trial~$\Delta$, the impurity self-energy equations are solved at every Matsubara frequency $\omega_n$ to obtain $\langle g_n\rangle_F$ and $\langle f_n\rangle_F$, which are summed to evaluate the subtracted gap equation residual; Brent's method then drives this residual to zero, yielding the self-consistent~$\Delta$. The supercurrent $J$ is computed from the converged Green functions via Eq.~\eqref{eq:current}.

For the spatial (inhomogeneous) problem, we start from initial profiles $\Delta(x) = \Delta_\text{BCS}(T)$ and $A(x) = -H_0\xi_0\exp(-x/\xi_0)$ and iterate: $A(x)$ is converted to the local Doppler shift $h(x) = \pi\Delta_0\xi_0\,A(x)$; the full local solve is performed at each grid point with the position-dependent $\gamma(x)$ and $h(x)$ to obtain $\Delta(x)$ and $J(x)$; $A(x)$ is then updated by solving the discretized Amp\`ere equation ${\rm d}^2A/{\rm d}x^2 = J(x)$ with the appropriate boundary conditions; and the old and new profiles are mixed with an adaptive relaxation parameter before checking for convergence. This sequence is repeated until the gap and vector potential profiles are mutually consistent to the prescribed tolerance.

For the uniform case, the solver returns the self-consistent gap~$\Delta$ and supercurrent density~$J$. For the spatial case, it returns the converged profiles $\Delta(x)$, $A(x)$, and $J(x)$ on the grid, from which derived quantities such as the magnetic field $B(x)={\rm d}A/{\rm d}x$ can be obtained.

The Fermi-surface averages are evaluated with Gauss--Legendre quadrature. After the substitution $u=\cos\theta$, the spherical Fermi-surface average becomes a standard integral on $[-1,1]$ with unit weight:
\begin{equation}
\langle{X}\rangle_F = \frac{1}{2} \int_0^\pi X(\theta) \sin(\theta) {\rm d}\theta = \frac{1}{2}\int_{-1}^1 X(u) {\rm d}u
\end{equation}
Because the Matsubara-axis integrands are smooth, unlike the corresponding real-axis integrands which have a discontinuity at the gap edge, only $N=16$ Gauss--Legendre nodes are needed to reach machine precision. A uniform trapezoid rule on $\theta$ would require $\sim200$ points for comparable accuracy.

We choose $T_c$ as the primary input material parameter. The gap equation itself is then rewritten in a UV-convergent subtracted form that eliminates the BCS coupling constant $g$ and Debye cutoff $\omega_D$ in favor of the clean-limit critical temperature $T_c$:
\begin{equation}
\ln\frac{T_c}{T} = \sum_{n=0}^{n_{\rm max}} \left[ \frac{1}{n+\tfrac{1}{2}} - \frac{\langle f_n \rangle_F}{\delta} \right], \qquad \delta \equiv \frac{\Delta}{2\pi k_B T}.
\end{equation}
The resulting summand decays as $\sim 1/n^3$, so a moderate energy cutoff $\omega_{\rm cut} \sim 30 \Delta_\text{BCS}(T=0)$ suffices without sensitivity to its precise value.

Rather than iterating the fixed-point equations $\tilde{\omega}_n = \omega_n + \gamma\langle g_n\rangle_F$, $\tilde{\Delta} = \Delta + \gamma\langle f_n\rangle_F$ to convergence, impurity self-consistency at each Matsubara frequency is cast as a two-dimensional root-finding problem for $(\langle g_n\rangle_F,\,\langle f_n\rangle_F)$. Two methods are tried in sequence, Powell hybrid followed by Levenberg-Marquardt, as implemented in SciPy \cite{2020SciPy-NMeth}, to handle possible convergence failures, and solutions from neighboring Matsubara frequencies are passed as initial guesses to accelerate convergence across the frequency sum.

The gap is likewise obtained from a one-dimensional root search: the nonlinear subtracted gap equation is reduced to a single scalar unknown $\delta = \Delta/(2\pi k_B T)$ and solved via Brent's method. At each trial~$\delta$, starting with the BCS gap, the full impurity self-consistency is resolved for all Matsubara frequencies. When no sign change is found in the bracketing interval, indicating the normal state, $\Delta=0$ is returned.

Amp\`ere's equation is handled as a banded linear solve. The second-order ordinary differential equation ${\rm d}^2A/{\rm d}x^2 = J(x)$ is discretized with a standard three-point first-order finite-difference stencil on a uniform grid, and the Neumann boundary condition ${\rm d}A/{\rm d}x = H_0$ at the surface is enforced with a second-order one-sided stencil $(-3A_0 + 4A_1 - A_2)/(2 x) = H_0$, avoiding the error that a first-order two-point stencil would introduce. The resulting tridiagonal-plus-one system, with lower bandwidth $l=1$ and upper bandwidth $u=2$, is solved with a banded LU factorization, as implemented in \texttt{scipy.linalg.solve}, at each outer iteration.

For the outer iteration between the gap/current solve and Amp\`ere's equation, adaptive under-relaxation is used: the mixing parameter~$\alpha$ is reduced down to a floor of $0.01$ when the error increases between iterations, and increased up to a ceiling of $0.7$ when the error decreases. Convergence is declared when the maximum change between iterations in both $\Delta(x)$ and $A(x)$, the latter converted to energy units via $h = \pi\Delta_0\xi_0 A$, falls below a specified tolerance, which can be specified, in our case, it was $1\cdot 10^{-6}$.

Throughout the calculation, all energy-like quantities ($T$, $h$, $\gamma$, $\Delta$, $\omega_n$) are carried in a single consistent energy unit, with no internal normalization by $\Delta_0$ or $T_c$; this avoids unit-conversion errors and allows the caller to work in any convenient system. The spatial grid and coherence length $\xi_0$ must likewise share a common length unit.

The code implementing this procedure is available online~\cite{GitHub-code}.

\section{Results} \label{sec.Results}
\subsection{$T_{c}$ enhancement with impurities}
We begin by examining how the critical temperature $T_c$ varies with the impurities for a constant impurity profile under external electromagnetic fields. 
To do so, we solve Eilenberger's equation to calculate the superconducting gap and see how it varies with temperature. 
$T_c$ is obtained when $\Delta=0$, meaning that there is a phase transition to the normal state.

\begin{figure}[h!]
    \centering
    \includegraphics[width=0.8\linewidth]{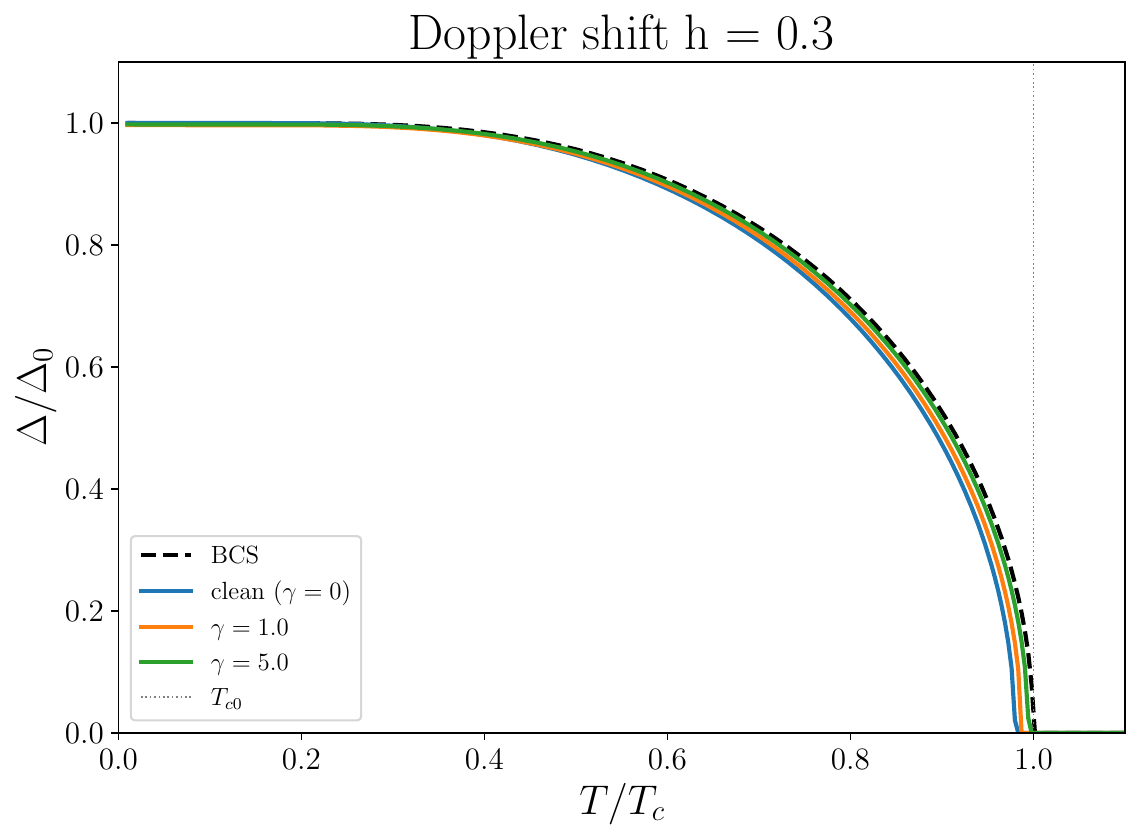}
    \includegraphics[width=0.8\linewidth]{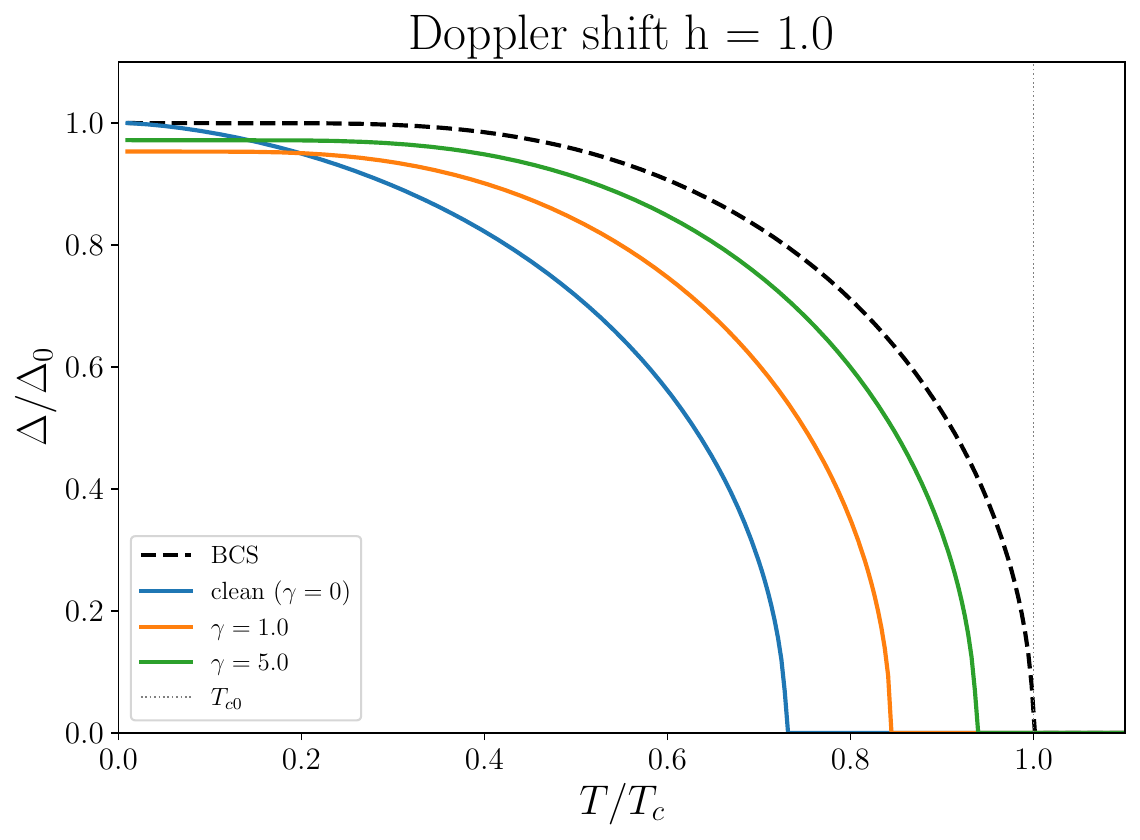}
    \caption{Critical Temperature enhancement by doping, where $h=(\hbar/2)|\bm{q}|$. As the vector potential is smaller, the effect of the impurities is reduced, in accordance with Anderson's theorem.}
    \label{fig:Gap_supression}
\end{figure}

Fig.~\ref{fig:Gap_supression} demonstrates the enhancement of $T_c$ by impurity scattering under an external field.
This can be explained because by adding impurities, the solution of Eilenberger's equation with impurities only differs from the one without by having an effective gap:
\begin{equation}
    \tilde{\Delta}= \Delta+\gamma\langle f\rangle_F,
\end{equation}
increasing the gap at most temperatures. 
This effect is more noticeable at higher fields that enhance the asymmetry of the Fermi surface, unlike the almost spherical Fermi surface at low fields, because $\hat{g}=\langle \hat{g}\rangle_F$ when the surface is symmetrical. Thus, the impurity term commutes with the Green's function matrix, indicating that no matter the amount of impurities added, they do not affect the superconductor in the absence of an external field, recovering Anderson's theorem, which implies that non-magnetic impurities in the absence of an external field do not affect the superconducting gap.

\begin{figure}[h!]
    \centering
    \includegraphics[width=0.85\linewidth]{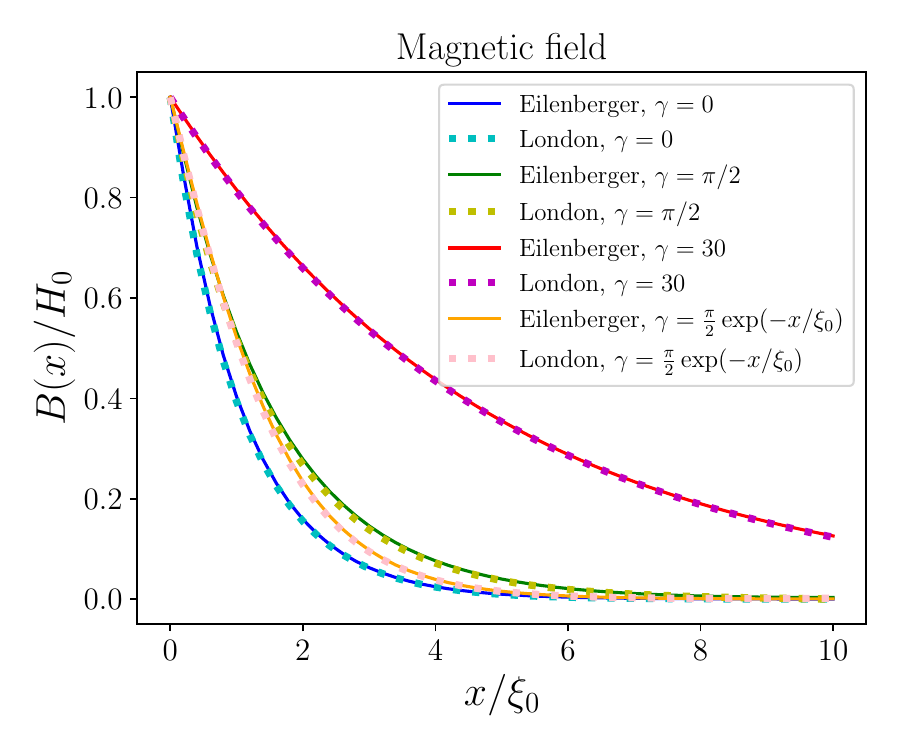}
    \includegraphics[width=0.85\linewidth]{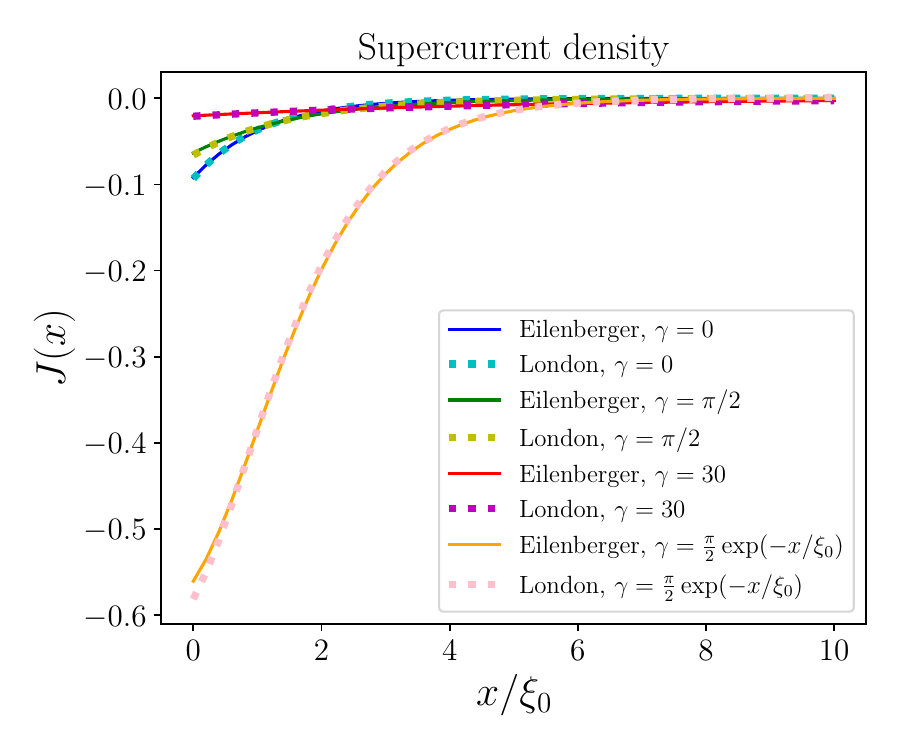}
    \caption{Magnetic field and supercurrent density profiles inside a superconductor for a constant impurity profile, in the clean limit ($\gamma=0$), the dirty limit ($\gamma=30$),  the crossover between clean and dirty limits ($\gamma = \pi/2$), and an exponential profile $\gamma(x)=\frac{\pi}{2}e^{-x/\xi_0}$. Solved for the Eilenberger and London equations at $T=0.5\cdot T_c$ and $H_0=0.1$. }
    \label{fig:meissner_profile_Constant}
\end{figure}

\subsection{Comparison between microscopic and macroscopic theories}  
We can also see how the magnetic field and superconducting current are affected by the impurities when the impurity profile is constant in Fig.~\ref{fig:meissner_profile_Constant}. This also serves as a check, since when the impurities are constant inside the superconductor, we recover the usual London equation, which is known to be valid.
Since there is no difference between the solutions in the constant profile case, it can be concluded that the Eilenberger solver works correctly. 
For the exponential impurity distribution $\gamma(x)$, which was studied in Ref.~\cite{PhysRevResearch.1.012015}, no differences between the two models can be found, meaning that the macroscopic theory with the modified London equation is a valid way of obtaining the magnetic profile inside a superconductor.
Up to now, Ref.~\cite{PhysRevResearch.1.012015} and Ref.~\cite{10.1063/5.0191234} are in principle identical.

\begin{figure}[h]
    \centering
    \includegraphics[width=0.9\linewidth]{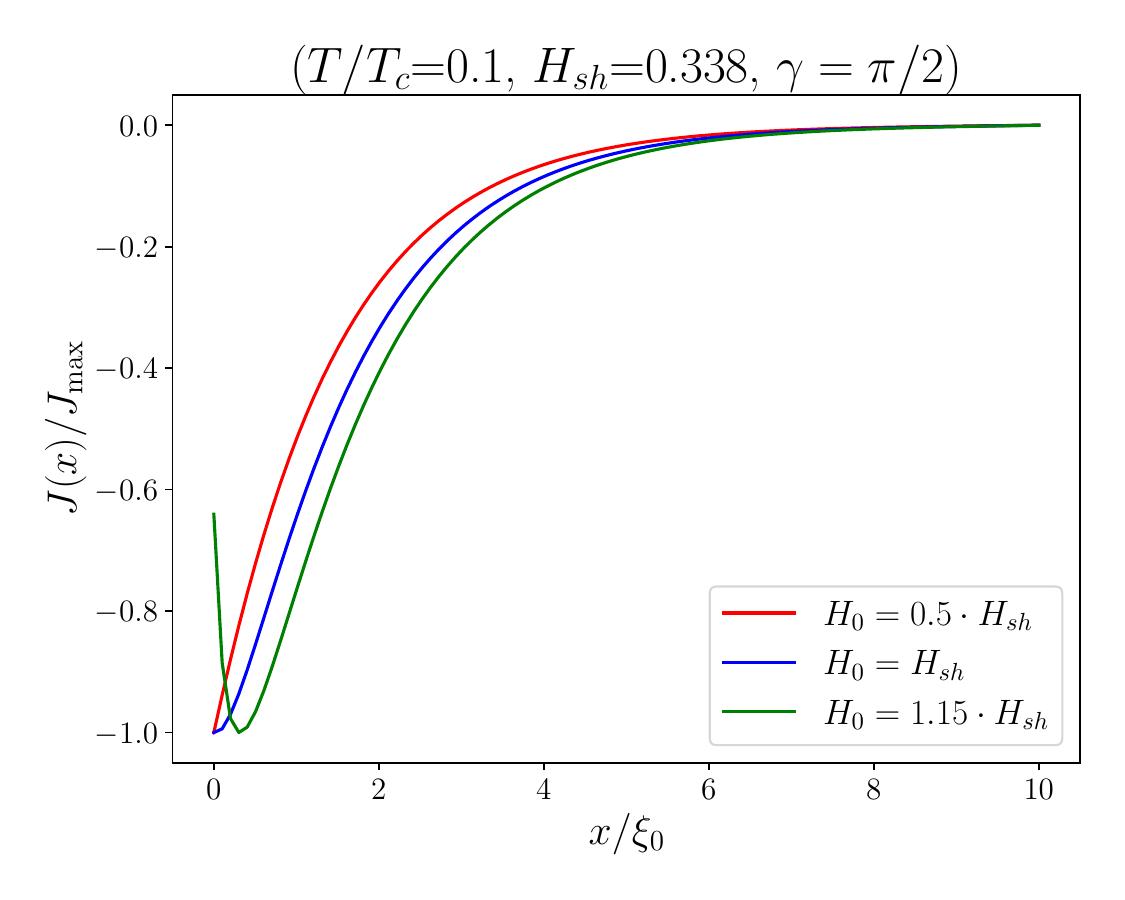}
    \caption{Change in the shape of the supercurrent density
as the external field approaches the superheating field. As the external field approaches the superheating field, the shape of the supercurrent density profile changes due to the transition to the mixed and normal states.}
    \label{fig:placeholder}
\end{figure}

The validity range of this approximated equivalence must be carefully examined, as mentioned in Sec.~\ref{sec.Theory}; the macroscopic approach only considers the penetration depth, resulting in a poor description near the superheating field as the phase transition to the mixed state occurs. 
As shown in Fig.~\ref{fig:placeholder}, the change in the shape of the supercurrent density clearly reflects this as the external field approaches the superheating field, and the superconductor undergoes a phase transition to the mixed state, and then the normal state.

Therefore, the London theory is valid for external fields lower than the superheating field. 
In the next subsection, a method to determine the superheating field is provided.

\subsection{$H_{ \rm sh}$ determination}
\begin{figure}[h!]
    \centering
    \includegraphics[width=0.9\linewidth]{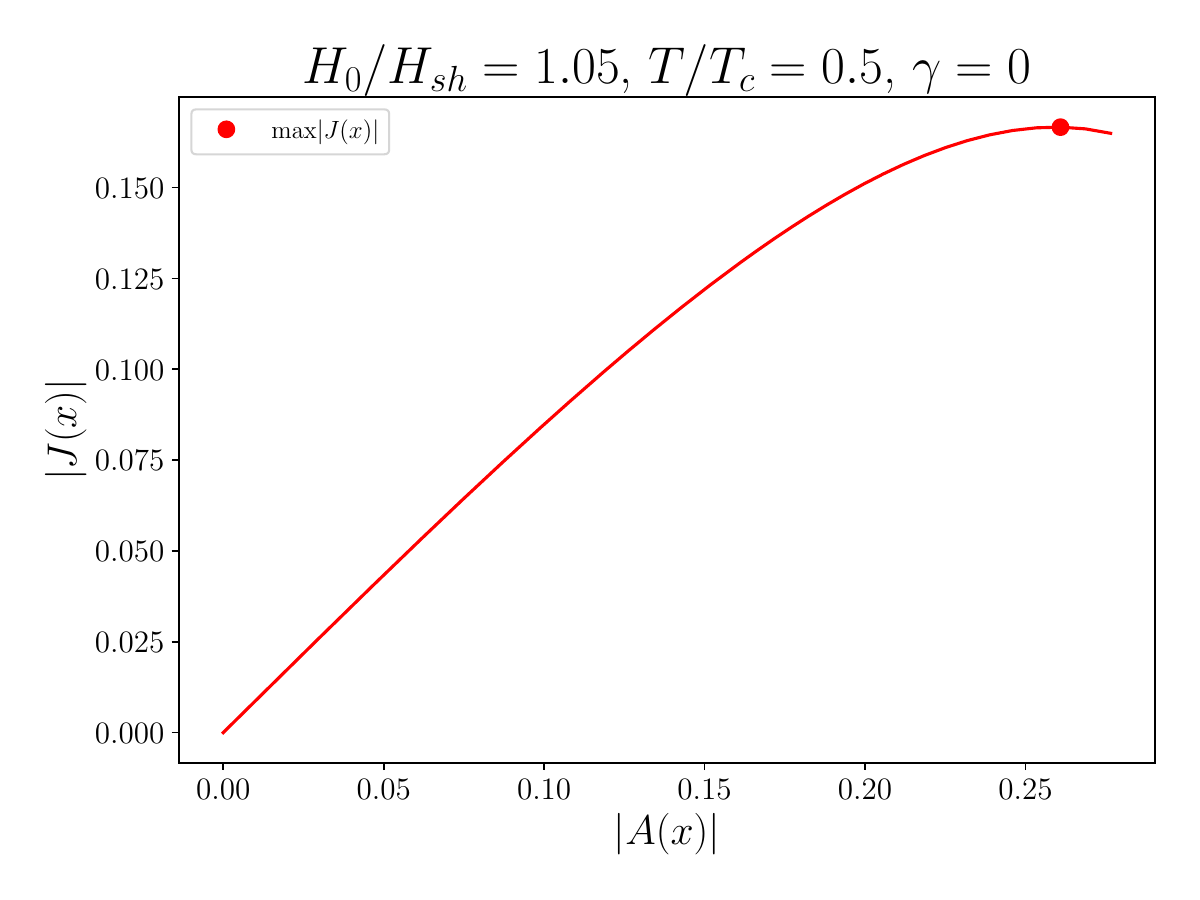}
    \caption{Superheating field computation by the described method. It is needed to know at which external field the magnitude of the supercurrent density is maximized. In fields higher than the superheating field, the Cooper pairs begin to depair due to the phase transition, making the supercurrent density decrease despite having a higher potential $\bm{A}$.}
    \label{fig:Superheating_Field_Determination}
\end{figure}

\begin{figure}
    \centering
    \includegraphics[width=0.9\linewidth]{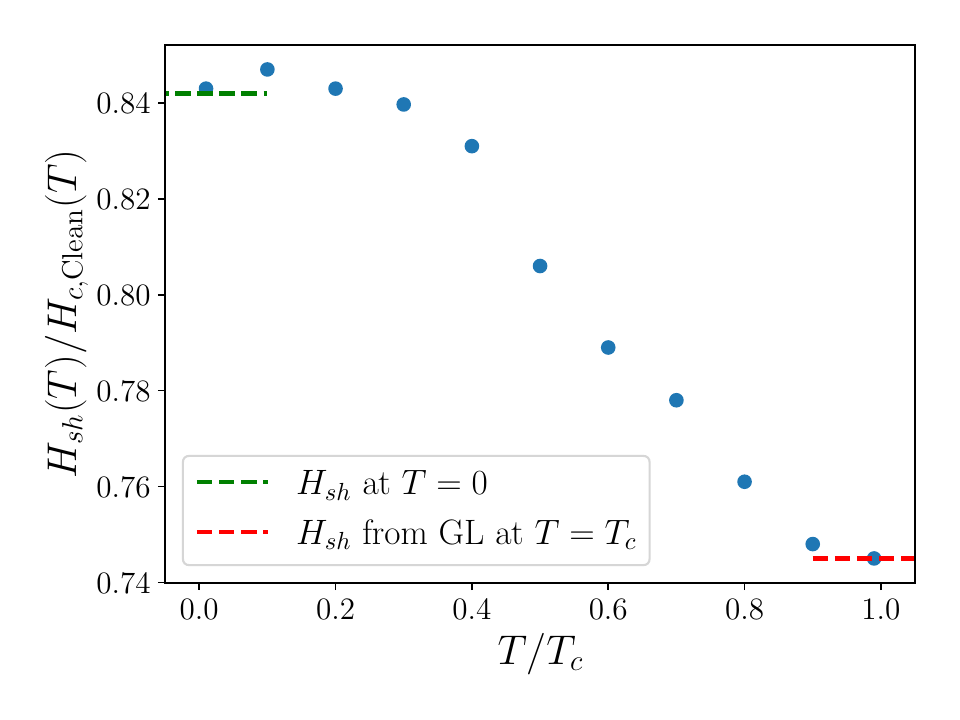}
    \caption{Superheating field in the clean limit ($\gamma=0$) for a type II superconductor. The results are compared with the known values at $T=0$ \cite{PhysRevB.78.224509} and $T=T_c$ \cite{PhysRevB.83.094505}.}
    \label{fig:Hsh_clean}
\end{figure}

As mentioned in the introduction, the superheating field is the field where the supercurrent density starts to be unstable. At higher fields, Cooper pairs will begin to depair, resulting in a lower superconducting current despite applying a higher external magnetic field.
Therefore, to compute the superheating field, we must solve Eilenberger's equation spatially and plot the supercurrent density against the magnetic vector potential and find which magnetic field produces the highest supercurrent, as demonstrated in Fig.~\ref{fig:Superheating_Field_Determination}.
Using this method, we can determine the superheating field at different temperatures in the clean limit, in the crossover region, and in the dirty limit.
As a sanity check of the theory, Figure~\ref{fig:Hsh_clean} compares the values of the superheating field produced by this method against those produced by the Ginzburg-Landau theory at $T=T_c$, and the known result in the literature for $T=0$ \cite{PhysRevB.78.224509}. 
Notably, since the phase transition occurs near the superheating field, the superconducting gap is sufficiently small for the Ginzburg-Landau theory to be a valid way to determine the superheating field at all temperatures~\cite{10.3389/femat.2023.1246016, PhysRevB.83.094505} except near absolute zero, as the Ginzburg-Landau theory is based on an effective action where the expansion parameter is $\Delta/T$.

\subsection{Optimal profile towards maximum $H_{ \rm sh}$ and macroscopic theory breakdown}
Before determining the best impurity distribution to maximize the superheating field, we need to consider the effect that impurities have on the field. 
Since the geometry is a semi-infinite one, the impurity effects are small at distances larger than the penetration depth. 
Also, since the superheating field is the field before vortex nucleation occurs, where vortices are pushed inside the superconductor by the supercurrent density~\cite{Kubo_2014,herrero2026analyticalevaluationsurfacebarrier}, if at a certain point we have zero supercurrent density, the superheating field should be highest. This is possible by having very localized impurities. 

\begin{figure}[h!]
    \centering
    \includegraphics[width=0.9\linewidth]{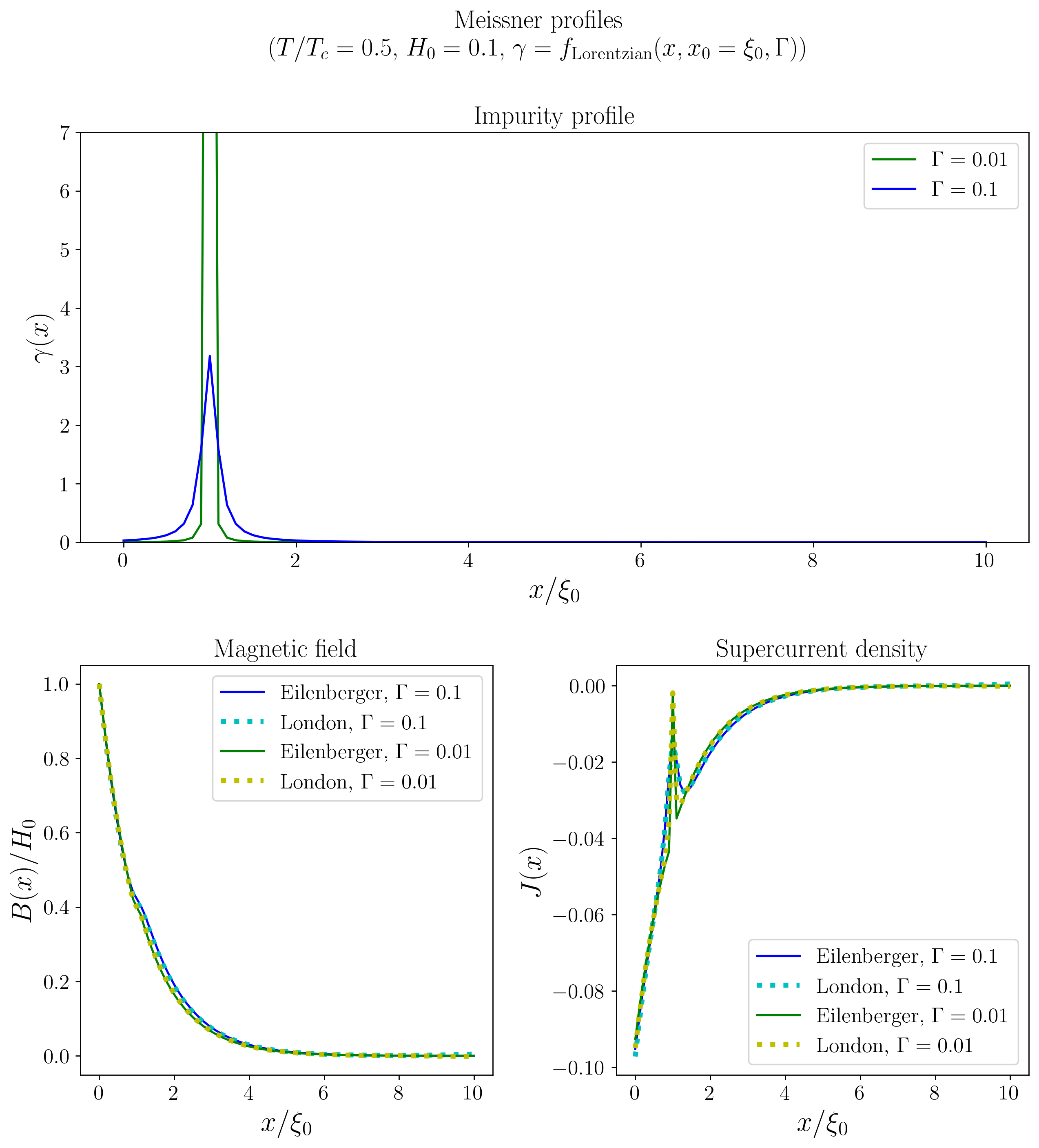}
    \caption{Magnetic field and supercurrent density inside a superconductor for a Lorentzian impurity profile. Solved for the London and Eilenberger equations.}
    \label{fig:Lorentzian}
\end{figure}

The localized impurities effectively act as an insulator, limiting the current at the point where they are implanted.
Finally, we study the optimal shape of the impurity profile by using a Lorentzian shape:
\begin{equation}
    \gamma_{\text{Lorentz}}(x,x_0,\Gamma)=\frac{1}{\pi}\frac{\Gamma}{(x-x_0)^2+\Gamma^2},
\end{equation}
and other possible shapes are a Gaussian or a family of functions that, at a certain limit, are a Dirac delta distribution.
Some typical calculations are shown in Fig.~\ref{fig:Lorentzian}.
Note that the macroscopic approach is still consistent with the microscopic computation up to now.

By computing the superheating field for nine different profiles based on this Lorentzian profile, where the peak and its width are changed, 
we identified the best set of parameters, as shown in Table~\ref{tab:placeholder}.
It can be seen that the best result is obtained for $\Gamma = 0.01,\ x_0=0$. 
This agrees with the expected results, as the current distribution protects the superconductor from vortices at the boundary, raising the superheating field and making it closer to the critical field. In the case where $\bm{J}=0$ at $x=0$, the superheating field coincides with the critical field; it should be noted that beyond a certain impurity level, there is no improvement since the supercurrent density would already become zero. 
This is also predicted by the macroscopic theory, as seen in Fig. \ref{fig:Lorentzian}.
\begin{table}[h!]
    \centering
     \refstepcounter{table}
    \textbf{Table \thetable: $H_{\rm sh}$ determination} \\[1ex]
    \begin{tabular}{|c|c|c|c|}\hline
         \diagbox{$x_0/\xi_0$}{$\Gamma/\xi_0$}&
  0.01& 0.1 & 1 \\ \hline
0 & 1.083 &1.055 & 0.911\\ \hline
0.25 & 0.855&  0.931     &0.896 \\ \hline
0.5 &0.844 &0.820 & 0.894\\ \hline
    \end{tabular}
    \caption*{Results of the superheating field $H_{\rm sh}(T)/H_{c,\text{Clean}}(T)$ for various Lorentzian profiles at $T/T_c=0.2$}
    \label{tab:placeholder}
\end{table}

An interesting observation is how, for $x_0 =0.25, 0.5,\ \Gamma= 0.1,0.01$ in Table~\ref{tab:placeholder}, the superheating field is lower than even the clean case. 
This is due to the definition of the superheating field, as there are barely any impurities before $x_0$, so that region of the superconductor is not protected against vortices entering. 
Therefore, vortices may enter the superconductor with a lower field, but they will be stopped at $x_0$, with the impurity profile effectively working as a Superconductor-Insulator-Superconductor device.
In this sense, the proposed impurity profile can be considered as a continuous limit of the authors' previous work about multilayer theory~\cite{herrero2026analyticalevaluationsurfacebarrier}.

\section{Summary and Outlook} \label{sec.Summary}
In this paper, we demonstrated the validity of a macroscopic theory to study a conventional superconductor with an inhomogeneous non-magnetic impurity profile. 
We also clarified the limitation of this approach near the superheating field $H_{\rm sh}$. 
We discussed which kind of impurity profile would be optimal based on vortex dynamics inside a superconductor.
Without a major surprise, the suggested distribution is a continuous limit of the multilayer theory to enhance the vortex penetration field.

Several future directions can be considered based on the results.
Without solving the microscopic theory, one can focus on the use of modified London equations justified by this paper.
This is encouraging for experimentalists when analyzing their experimental data.
The proposed optimum distribution of the non-magnetic impurities could be engineered in a real material.
The most natural way would be through some layer deposition techniques.
We focus on the non-magnetic impurities expressed as the self-energy $\Sigma_{\rm imp}$.
Several groups have recently been making efforts to include magnetic impurities in the optical conductivity of superconductors~\cite{PhysRevB.94.144508,zarea2026}.
Extending our results to magnetic impurities represents a promising future direction.

\begin{acknowledgments}
We would like to thank A. Perez Ruiz, C. Cerna, H. Hu, and D. Bafia for useful discussions.
We would also like to acknowledge discussions with L. Wallace at TTC 2026, who is studying the same problem using Ginzburg-Landau formalism, as well as A.J. Koren, who did a thorough proofreading and pointed out some mistakes. 
AM and CRH are also grateful towards ANL and FNAL hospitality, as part of this work was carried out at these institutions, as well as J.A Sauls' group at LSU for discussing related topic during a short stay there. 
This work was supported by the CNRS-UChicago IRC grant, FACCTS funding at UChicago, and the European Union’s Horizon Europe research and innovation programme under Grant Agreement No. 101086276 (EAJADE).
\end{acknowledgments}


\bibliography{apssamp}

\end{document}